\documentclass[aps,floats,prx,showpacs,twocolumn]{revtex4-1}

\usepackage{graphicx,subfigure}
\usepackage{amsfonts,amsmath} \usepackage{bm} \usepackage{dcolumn}
\usepackage{epsfig} \usepackage{latexsym}
\usepackage{graphicx}
\usepackage{multirow}
\usepackage{color}

\begin{document}

\title{Intertwined order and holography: the case of the parity breaking pair density wave}

\author{Rong-Gen Cai$^{a}$}

\author{Li Li$^{a,b}$}

\author{Yong-Qiang Wang$^{c}$}

\author{Jan Zaanen$^{d}$}

\affiliation{$^a$ CAS Key Laboratory of Theoretical Physics, Institute of Theoretical Physics, \\
Chinese Academy of Sciences, Beijing 100190, China\\
 $^b$ Department of Physics, Lehigh University, Bethlehem, PA, 18018, USA\\
$^c$ Institute of Theoretical Physics, Lanzhou University, Lanzhou 730000, China\\
$^{d}$ Institute Lorentz for Theoretical Physics, Leiden University, Leiden, The Netherlands}


\begin{abstract}

We present a minimal bottom-up extension of the Chern-Simons bulk action for holographic translational symmetry breaking that naturally gives rise to
pair density waves.  We construct stationary inhomogeneous black hole solutions in which both the U(1) symmetry and spatially translational symmetry are spontaneously broken at finite temperature and charge density. This novel solution provides a dual description of a superconducting phase intertwined with charge, current and parity orders.
\end{abstract}


\maketitle

{\em Introduction.}\;\;\;
The explanation of states of matter that break symmetry spontaneously in circumstances where quantum effects are dominating has been 
a traditional mainstay of condensed matter physics. Famous examples are the BCS theory of conventional superconductivity as well as the 
Peierls theory explaining charge and spin density waves in terms of the nesting of the Fermi surface. 
As an outcome of a long empirical development it was discovered that the strongly interacting systems realized in underdoped cuprate superconductors  seem to realize a remarkably complex form of ``intertwined" orders~\cite{naturerev15,fradkinPDW}.  There are indications for exotic, quantum physical forms of order that are not found elsewhere. A first example is the pattern of {\em spontaneous electronic ``orbital" currents}  encircling the plaquettes of the copper oxide planes~\cite{currentsfirst,currentsGreven,currenttflucGreven}, while there is definitive evidence for the breaking of parity~\cite{parityodd}. Another example is the {\em pair density wave} (PDW): a superconducting state that does break translations in zero magnetic field, for which experimental evidence was  reported in spin-striped 214 superconductors~\cite{tranquadaPDW07,Rajasekaran:2017} and very recently in a charge ordered 2212 ``BISCO" superconductor~\cite{DavisPDW16}. Departing from the established repertoire of condensed matter theories it appears to be quite difficult  to explain the origin for the observed patterns of symmetry breaking in cuprates~\cite{Devstripes,Simonsnum}. 
The details of the conventional symmetry breaking (like the periodicity of the charge order) are far from being understood~\cite{naturerev15}, 
while a convincing explanation of the mechanisms leading to spontaneous currents and PDW's is still lacking. Last but not least, it appears that these orders occur simultaneously in a specially orchestrated ``intertwined"  relationship, for reasons that are largely mysterious~\cite{fradkinPDW}.

Holographic duality has a track record as a useful theoretical laboratory to explore the rich structure of quantum matter characterized by densely entangled degrees of freedom~\cite{thebook15,Hartnoll:2016apf}. It maps the quantum problem in the ``boundary" onto classical gravitational physics in a
``bulk" space-time with one extra dimension. The rules of effective field theory (EFT) based on symmetry and locality are applied to the bulk, and this dualizes in a set of robust and universal phenomenological theories describing the physics in the boundary. The application of holography to condensed matter jump-started by the discovery of holographic superconductivity~\cite{Hartnoll:2008vx} showing that spontaneous symmetry breaking has a stunningly elegant gravitational dual~\cite{thebook15}: the black holes carry a halo of charged hair which spontaneously breaks a global U(1) symmetry of the dual field theory. 
This shares similarities with BCS superconductivity in the boundary~\cite{thebook15}, but yet quite different~\cite{She:2011cm} given that the 
normal state is a holographic {\em strange metal}~\cite{thebook15} instead of a Fermi-liquid. We will demonstrate here that by further developing the bulk theory using the EFT rule book, a pattern of intertwined order emerges in the boundary sharing intriguing similarities with those realised in the cuprates. This suggests that the peculiar intertwined nature of the low temperature order may be rooted in the densely entangled nature of the quantum critical metal. The next ingredient is bulk {\em topological} terms -- Chern-Simons (CS)~\cite{Nakamura:2009tf} and theta terms~\cite{Donos:2011bh} in odd and even dimensions, respectively. Intriguingly, these describe the emergence of spontaneous {\em current order intertwined with the crystallisation of charge}.

\begin{figure}[ht!]
\begin{center}
\subfigure[]{\includegraphics[scale=0.33]{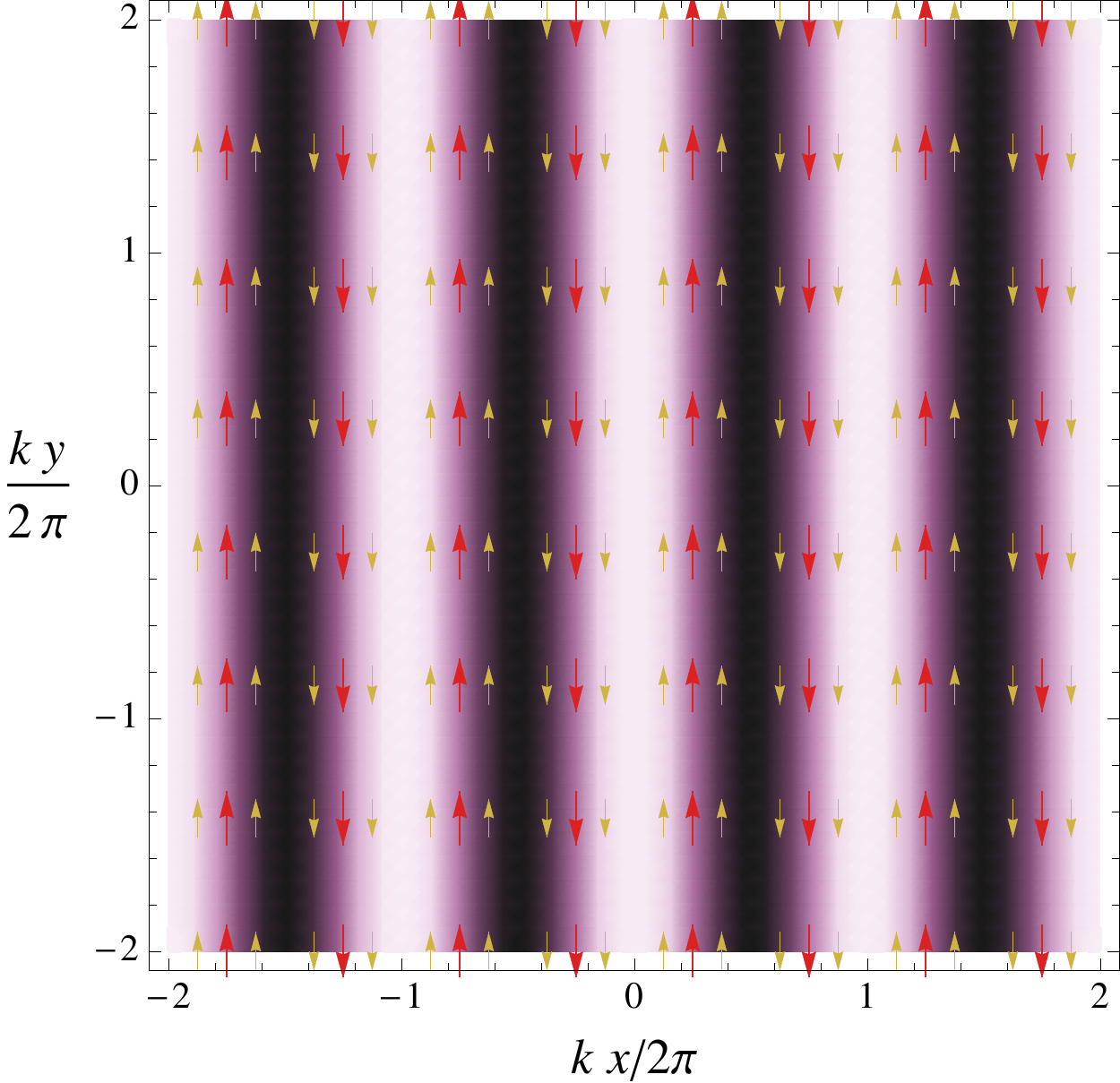}}
\subfigure[]{\includegraphics[scale=0.33]{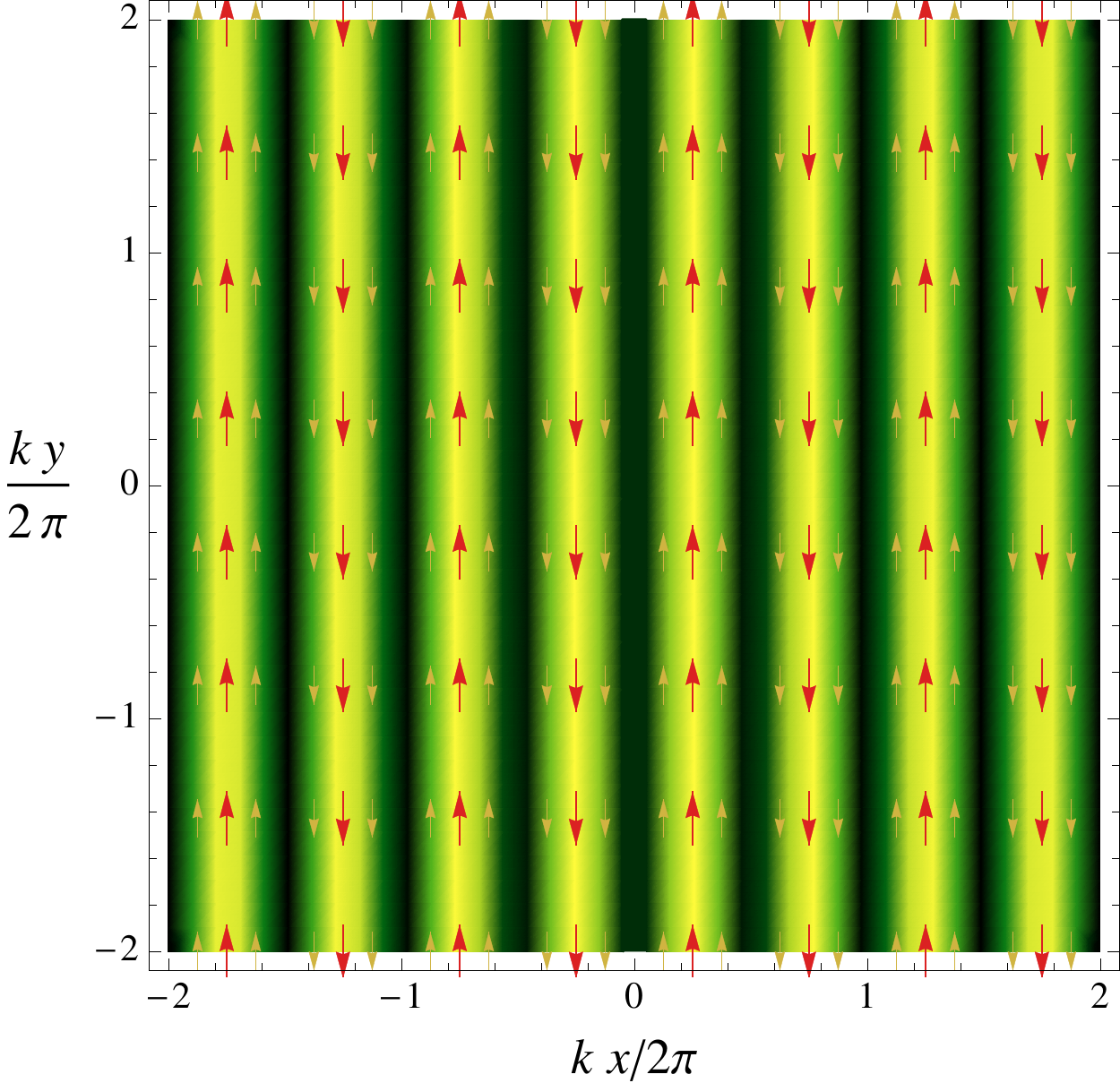}}
\subfigure[]{\includegraphics[angle=0,scale=0.22]{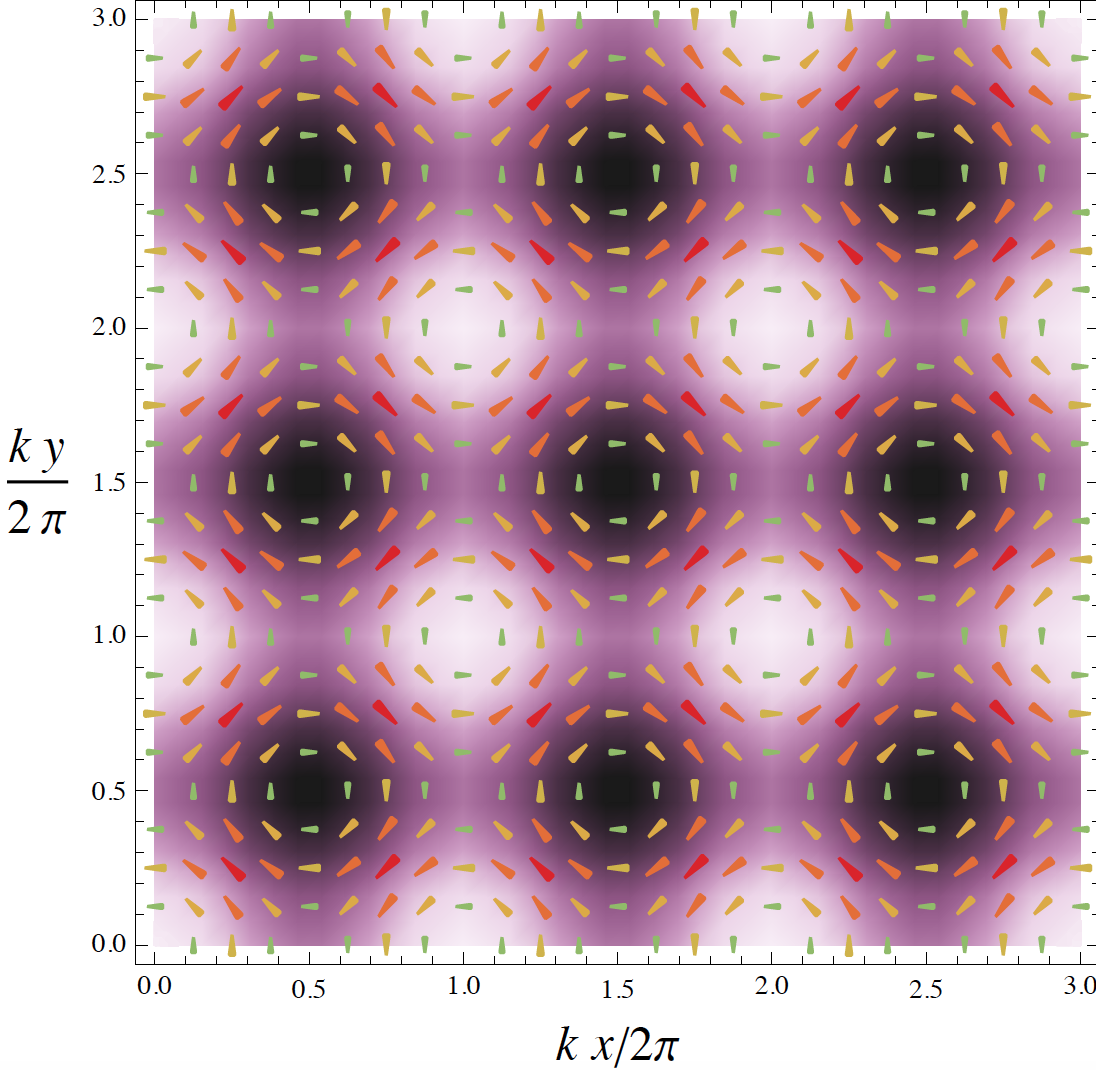}}
\subfigure[]{\includegraphics[angle=0,scale=0.22]{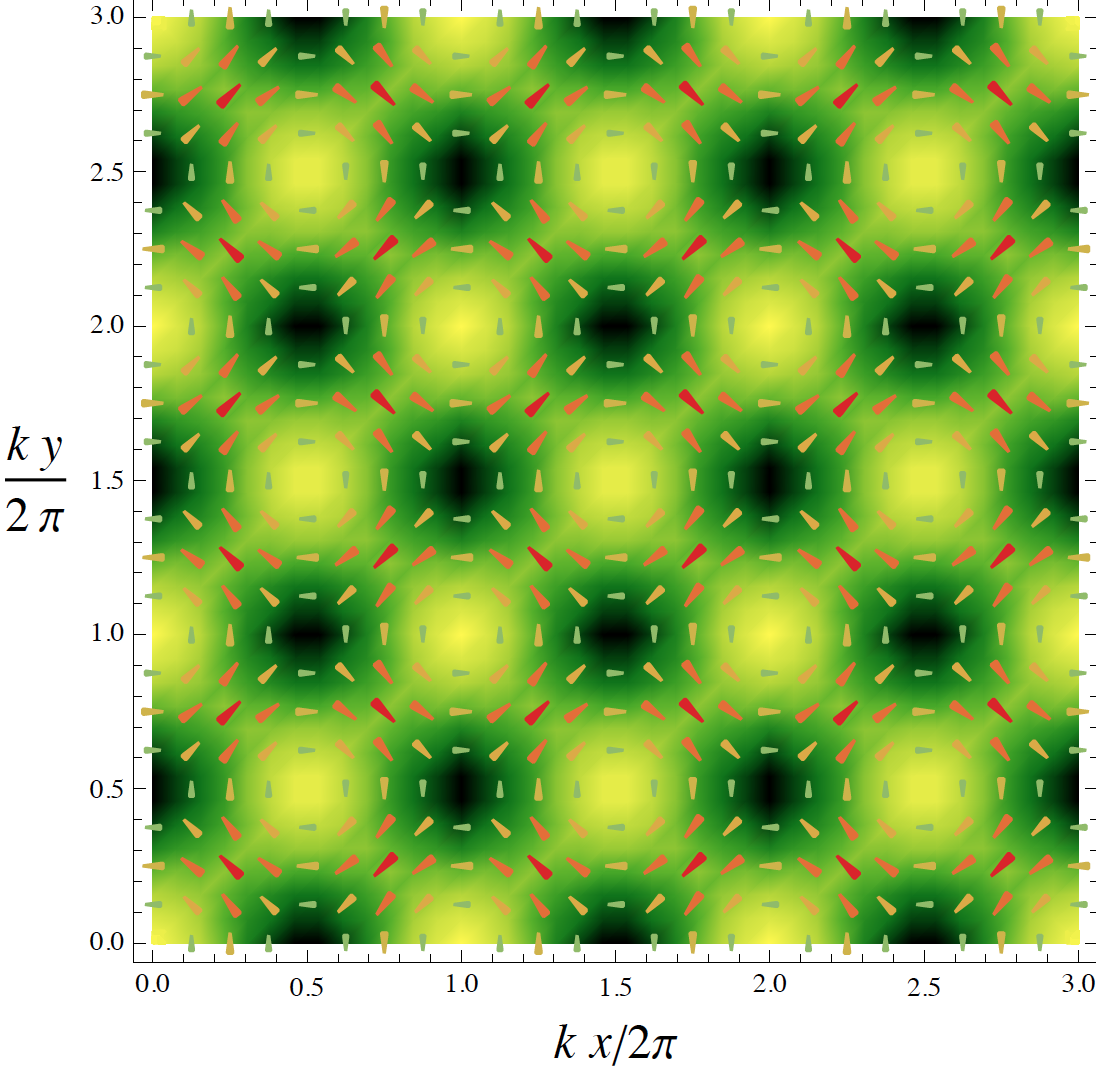}}
\caption{The density plots of (a) condensate and (b) charge density distribution for uni-directional phase at $T/T_c\approx0.64$, and that of (c) condensate and (d) charge density distribution for the square checkerboard at $T/T_c\approx0.94$. The arrows denote integral curves of the currents $(\left<J^x\right>,\left<J^y\right>)$. The bright parts correspond to large positive values while the dark region to the negative values. In the uni-directional phase the charge density oscillates at twice the frequency of the currents and condensate. The latter two orders are precisely out-of phase. The checkerboard is a tetragonal charge crystal going hand in hand with a staggered pattern of current fluxes circling around the plaquettes, which is similar to the d-density waves of condensed matter physics.}
\label{fig:density}
\end{center}
\end{figure}

We present here a minimal extension of the CS type bulk theory to incorporate holographic superconductivity. The outcome is that the superconductivity generically turns into a pair density wave, automatically implying the simultaneous spontaneous breaking of parity~\cite{parityodd}. Representative results are shown in Fig.\,\ref{fig:density} for a uni-directional and a ``tetragonal" crystallization in two space dimensions. The rules governing these patterns are most easily inferred from the uni-directional case Fig.\,\ref{fig:density}(a,b). The charge accumulates in ``stripes" (bright areas panel (b)) where the current densities are maximal. The currents run in the perpendicular $y$ direction and these form a staggered pattern with the maxima being coincident with the maximal charge density, vanishing in the middle of the low density domains. The superfluid density (panel (a)) exhibits the same periodicity as the current order, but it is precisely out-of phase with the latter: it has nodes where the current and charge density are maximal while it oscillates from maximal negative to positive values in the low density domains. This is precisely what is meant with a pair density wave. These rules repeats themselves in the tetragonally (``checkerboard") ordered case. The charge order (panel (d))  is now accompanied by spontaneous staggered current patterns~\cite{Withers:2014sja} similar to  the  ``d-density wave"~\cite{Nayak, Chakravarty,affleck,Kotliar}, having quite a history in the condensed matter literature. This is now accompanied with a 2D PDW having twice the periodicity of the charge order, changing sign from one ``interstitial" region to the next where it acquires a maximum absolute value.  Let's now discuss in more detail how we arrive at these results.

{\em Gravity setup.}\;\;\; 
Our starting point is a (3+1) dimensional gravity theory  dual to the (2+1) dimensional boundary theory containing a gauge field $A_\mu$ and two scalar fields $\chi$ and $\theta$, assuming the coordinate system $(t, x, y, z)$ in which the AdS boundary is located at $z=0$:

\begin{eqnarray}\label{action}
S=\int d^{4}x&& \sqrt{-g}\Big [\mathcal{R}+6-\frac{1}{2}\partial_{\mu}\chi \partial^{\mu}\chi-\mathcal{F}(\chi)(\partial_\mu\theta- A_\mu)^2\nonumber \\
-&&\frac{Z(\chi)}{4}F^2-V(\chi)-\vartheta(\chi) \epsilon_{\mu\nu\lambda\sigma}F^{\mu\nu}F^{\lambda\sigma}\Big]\,.
\end{eqnarray}

Ignoring the fourth term that involves $\theta$ this reveals the canonical holographic  mechanism for spontaneous translational symmetry breaking. One recognises the theta term $\sim\chi F\wedge F$ (with  
$\epsilon_{txyz}=\sqrt{-g}$)
as inspired on top-down holography by consistent truncation of 11D supergravity~\cite{Gauntlett:2009bh}, 
 while fully back-reacted geometries can be obtained~\cite{Rozali:2012es,Withers:2013kva,Donos:2013wia,Donos:2015eew}. 
 Viewed from a bottom-up perspective, such a topological term is generic and it can be added to the bulk in order to respect general symmetry principles.  The topological term has the effect to shift the instability (BF bound violation in the bulk) to a finite wave vector, but for this to happen one needs simultaneous  ``intertwined" VEV's of the currents and the charge density. This is responsible for the intertwined charge-current order in Fig.\,\ref{fig:density}. We notice that in any other holographic setup the breaking of translations involves meticulous fine-tuning, {\em e.g.}~\cite{Donos:2013gda} and in this sense the intertwinement of charge and current orders follows naturally from holography.
 
Our novelty is to include the $\theta$ field, that represents an easy way to include the minimal coupling of $\chi$ to the gauge field $A_\mu$  by a St\"uckelberg  term (known as ``Josephson action" in the condensed matter literature). This has the advantage of admitting less restrictive scalar couplings: in this generalized class of theories the two real degrees of freedom $\chi$ and $\theta$ are not necessarily associated with the magnitude and phase of a complex scalar, respectively. 
Holographic superconductors in similar setups have been studied for instance in Refs.~\cite{Franco:2009yz,Aprile:2009ai,Cai:2012es,Cremonini:2016rbd,Gouteraux:2016arz}. The bulk solution with non-trivial $(\chi,\theta)$ corresponds to the U(1) broken phase. Although there is no top-down justification present available, the structure of this action suggests a minimal way to merge the superconductivity and the flux phase. As $\theta$ can be absorbed by gauge fixing, the condensation of the scalar operator dual to $\chi$ could play the role of superconducting order parameter. The bulk St\"uckelberg term implies the breaking of U(1) in the boundary and the presence of the superconducting condensate is revealed by the characteristic delta function 
at zero frequency in the optical conductivity (see supplementary material~\cite{addsm}).
   
In this letter we focus on the particular model with 
\begin{equation}\label{couplings}
\begin{split}
&Z(\chi)=\frac{1}{\cosh(\sqrt{3}\chi)}\,,\quad \quad V(\chi)=1-\cosh(\sqrt{2}\chi)\,,\\
&\mathcal{F}(\chi)=\cosh(\chi)-1\,,\quad \vartheta(\chi)=\frac{1}{4\sqrt{3}}\tanh(\sqrt{3}\chi)\,.
\end{split}
\end{equation}
The scalar operator $\mathcal{O}$ dual to $\chi$ has therefore dimension $2$. The normal phase with $\chi=0$ is described by 
the AdS Reissner-Nordstr\"{o}m (AdS-RN) black brane~\cite{Cai:1996eg},
\begin{eqnarray}\label{RNads}
ds^2&&=\frac{1}{z^2}\left[- H(z)dt^2+\frac{1}{H(z)}dz^2+(dx^2+dy^2)\right]\,,\\
H(z)&&=(1-z)\left(1+z+z^2-\frac{\mu^2 z^3}{4}\right)\,,\; A_t=\mu(1-z)\,,\nonumber
\end{eqnarray}
where $\mu$ is the chemical potential, rescaling the coordinates so that the horizon is at $z=1$. The temperature is given by $T/\mu=(12-\mu^2)/16\pi\mu$. Departing from the extremal RN  $AdS_2\times \mathbb{R}^2$ background it follows from perturbative analysis that a homogeneous  U(1) broken solution is excluded for the model~\eqref{couplings}~\cite{CLWZ}. The last term plays a crucial role, imposing that the breaking of the U(1) symmetry goes hand-in-hand with the breaking of translational invariance. It follows immediately from this bulk action that parity is broken as well in the simultaneous presence of the superconducting, charge and current orders with a VEV proportional to the U(1) order parameter~\cite{addsm}.

{\em Intertwined Black holes.}\;\;\;
Let us now construct the inhomogeneous intertwined black holes, focussing first on the uni-directional ``striped" solutions. The procedure  leading  to the fully crystallised solutions is just a straightforward generalisation involving two spatial directions. 
We seek for uni-directional solutions within the following ansatz,
\begin{eqnarray}\label{ansatz}
ds^2=&&\frac{1}{z^2}\Big[-H(z) U_1 dt^2+\frac{U_2}{H(z)}dz^2+U_3(d x+z^2 U_5 dz)^2 \nonumber \\
&&\quad\quad\quad\quad\quad+U_4(dy+(1-z)U_6 dt)^2\Big]\,,\\
A=&&A_t \,dt+A_y\, dy\,,\quad \chi=z\, \psi\,,\nonumber
\end{eqnarray}
with nine functions $(\psi, A_t, A_y, U_i), i=1,...,6$, depending on $z$ and the spatial coordinate $x$ involved in the symmetry breaking, while $H(z)$ is given in eq.~\eqref{RNads}. This results in a system of equations of motion involving 9 PDE's in terms of the variables $z$ and $x$. We employ the DeTurck method~\cite{Headrick:2009pv} to solve the system. For the tetragonal case we deal with 15 functions that all depend on the radial coordinate $z$ and two spatial coordinates $(x,y)$. To solve such a set of cohomogeneity-three PDE's is challenging due to a dramatic increase in computational complexity.

Spontaneous symmetry breaking means that all sources should be turned off, while fixing the  boundary metric to be asymptotically AdS at the UV boundary $z=0$,
\begin{eqnarray}
&&U_1(0,x)=U_2(0,x)=U_3(0,x)=U_4(0,x)=1\,,\\
&&\psi(0,x)=A_y(0,x)=U_5(0,x)=U_6(0,x)=0,\, A_t(0,x)=\mu\,.\nonumber
\end{eqnarray}
We impose the regularity condition at the horizon $z=1$ and therefore all the functions have an analytic expansion in powers of $(1-z)$. On the spatial boundary one imposes a periodicity condition. With those boundary conditions, the set of inhomogeneous solutions is parametrized in terms of the temperature $T/\mu$ and wave-number $k/\mu$, determining  the periodicity in the symmetry broken $x$ direction. 
\begin{figure}[ht!]
\begin{center}
\subfigure[]{\includegraphics[scale=0.36]{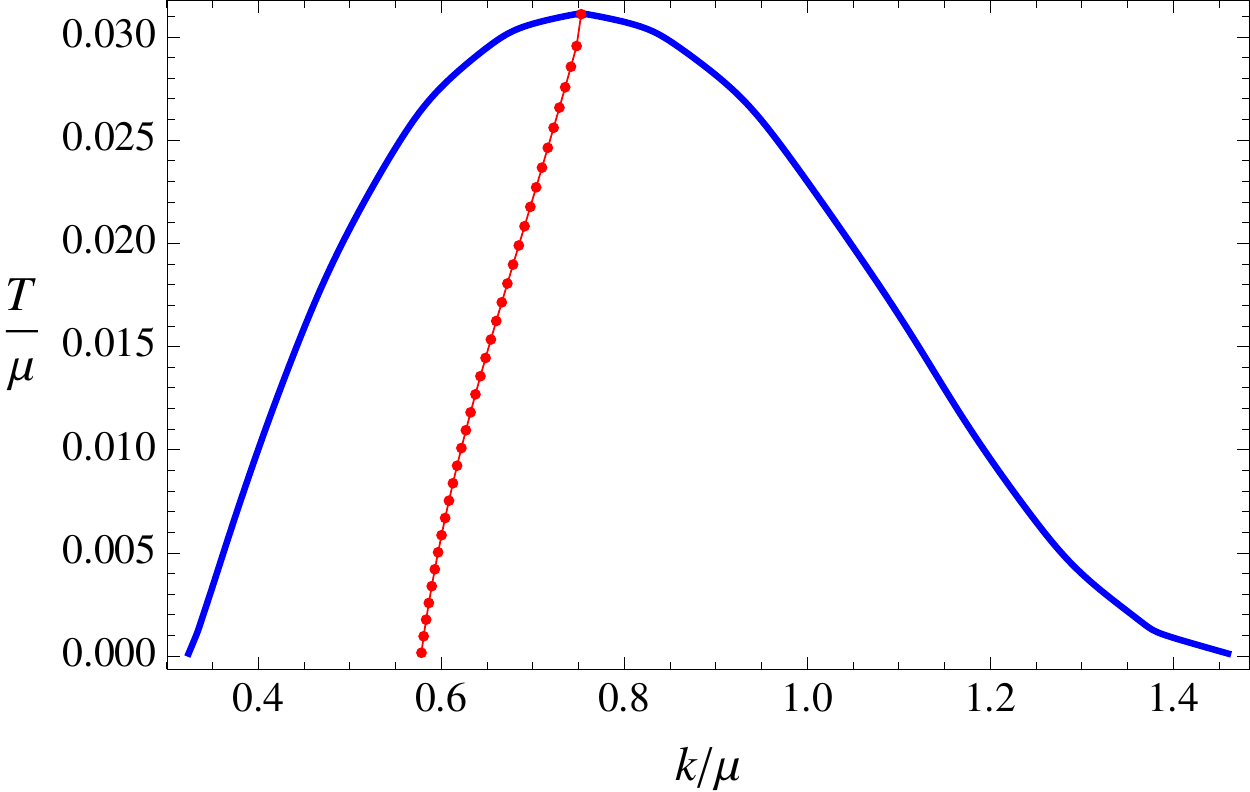}}
\subfigure[]{\includegraphics[scale=0.55]{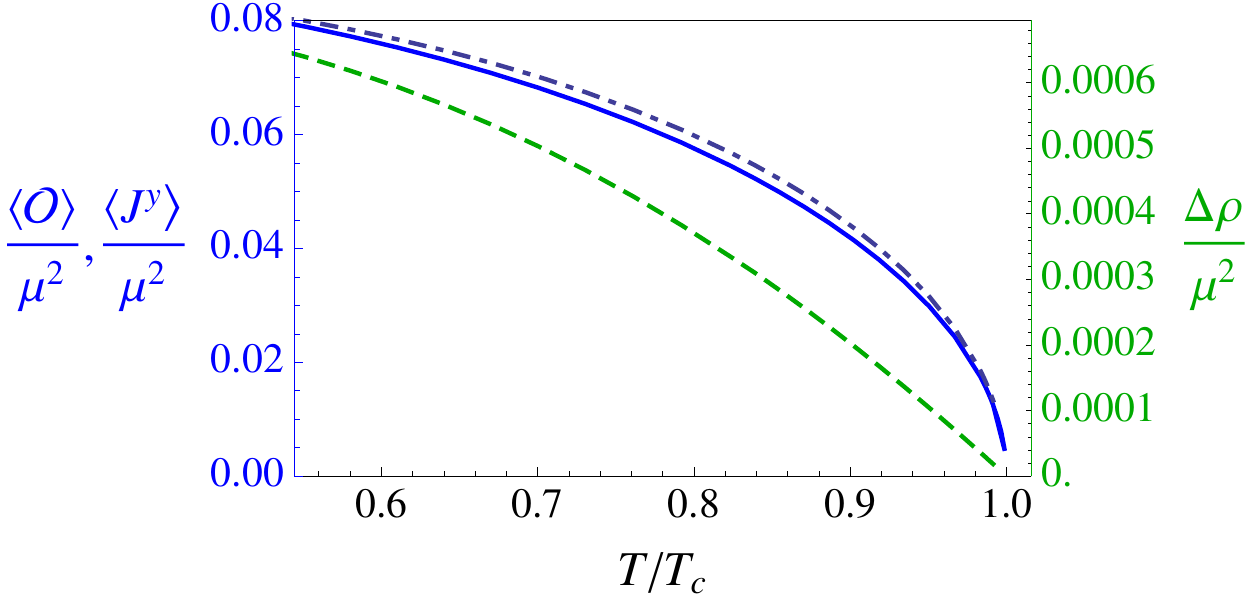}}
\caption{ (a) The stability ``dome" (blue line) following from linear stability analysis in the  temperature ($T/\mu$) -- ordering wave vector ($k /\mu$) plane ($\mu$ is the chemical potential) for the unidirectional 
order. The actual second order transition is associated with the maximal $T_c$ and the red dotted line shows the temperature dependence of the ordering wave vector of the fully back-reacted ``intertwined" order. 
(b) The temperature dependence of the charge modulation $\Delta \rho$ (dashed), current $\langle J^y \rangle$ (dash-dotted) and pair density wave $\langle O\rangle$ (full line) VEV's of the uni-directional phase. The dominating current and pair density wave VEV's show a characteristic mean field temperature evolution, while the ``parasitic" charge density modulation $\Delta \rho$ increases linearly. Non-trivial configurations of $\langle O\rangle$ also imply a spontaneous breaking of parity, which was very recently observed in the cuprates~\cite{parityodd}. Similar results are obtained for the full tetragonal order of Fig.\,\ref{fig:density}.}
\label{fig:stripeTk}
\end{center}
\end{figure}

We show the critical temperature as a function of wave-number in Fig.\,\ref{fig:stripeTk}(a) derived from linear stability analysis of the normal state. This curve is peaked at a non-zero value $k_c\approx 0.75\mu$ with a $T_c\approx0.03\mu$: this is where the continuous thermal phase transition occurs from the normal system to a uni-directional striped phase with periodicity $k_c$.  
At temperatures below $T_c$ non-linearities become important and the full set of 9 (or 15 for tetragonal case) coupled PDE's has to be solved. We have numerically constructed  the solutions parameterized in terms of $T/\mu$ and $k/\mu$, finding that the free energy is minimized at a weakly temperature dependent ordering wave vector~\cite{Rozali:2012es,Withers:2013kva} indicated by the red dotted line in Fig.\,\ref{fig:stripeTk}(a) for the  uni-directional case, with similar results for the tetragonal case~\cite{CLWZ}.

In Fig.\,\ref{fig:stripeTk}(b) we show the temperature dependence of the various order parameters as a function of temperature. The current and PDW order parameters  have a similar temperature dependence indicative of a Landau mean-field second order transition $\sim\sqrt{T_c -T}$. The charge order grows more slowly $\sim (T_c -T)$; this is indicative of the current order/PDW being the {\em dominant} order parameter while the charge modulation is induced parasitically. This is a generic affair which is well  understood in terms of Landau theory as  of relevance also to the ordering of spin-stripes~\cite{Zachar}.  The lowest 
 order invariant coupling the order parameters is $\rho |\langle O \rangle |^2$ ($\rho |\langle J^y \rangle |^2$). When the order occurs at finite wave-number this immediately implies that the period of the charge order is twice that of the current/PDW order. However, it also implies that when the dominant (mean-field) instability is associated with the currents and the  PDW the charge density modulation will grow linearly with temperature below $T_c$.  Interestingly, this is accomplished in the bulk by the leading non-linear correction near $T_c$. The current and condensate modes are in combination violating the BF bound, and drive the instability. In linear order the charge density is decoupled but in next-to-leading order the modulated $A_y$ and $\chi$ fields induce a charge density modulation growing linearly with temperature.  Eventually, at low temperatures non-linearities encapsulated by the fully back-reacted bulk solutions become quite important giving rise to the intertwined ordering pattern shown in Fig.\,\ref{fig:density} which we already discussed.

{\em Discussion.}\;\;\;
There is obviously still a long way to go in  order to address the ordering patterns realized in underdoped cuprates in any detail. At the present stage the outcomes of the holographic exercise presented in the above offer no more than a rough cartoon. However, the cartoon is suggestive with regard to generalities. Departing from a strange metal normal state, holographic ``naturalness" seems to insist that the spontaneous symmetry breaking at low temperature should necessarily give rise to intertwined order involving besides charge order also spontaneous currents, parity breaking and the pair density waves. This rests on applying the rules of EFT to the gravitation dual: the simplest way to break translations is by invoking the topological (theta and CS) terms that automatically intertwine  charge and current orders. We have presented here the most minimal gravitational theory that resurrects the superconductivity, turning it automatically in a pair density wave, remarkably at the ``expense" of an unavoidable spontaneous breaking of parity.

Is there room to make this more realistic? There are some intrinsic limitations. The temperature dependence of order parameters (Fig.\,\ref{fig:stripeTk}) should be taken with a grain of salt. Because of the matrix large N limit associated with holography thermal fluctuations are completely suppressed, while it is well established that these are important both for the PDW~\cite{tranquadaPDW07,Rajasekaran:2017} and the current order~\cite{Varma,Shekhter}. A crucial missing ingredient is the periodic background potential. The present construction lives in the Galilean continuum while in experiment the ionic background potential is playing an important role. This complicates the bulk physics further and only very recently first results became available showing commensurate lock-in between spontaneous translational order and a background periodic potential~\cite{Andrade:2017leb}. 
There is a wealth of phenomena to explore. Is it possible to construct holographically the current-loop order living inside the $CuO_2$ unit cells by applying an appropriate background potential? 
It has been all along clear that the cuprate stripes/CDW are eventually rooted in discommensuration effects associated with the competition between the periodicity of the background and spontaneous ``lattices"~\cite{ZaGu,mesarosSTS}.  

Another major limitation of state of the art holography is related to the truly relativistic nature of the boundary field theory. The matter described in the boundary is formed from massless degrees of freedom with the ramification that spin and orbital motions are locked together in helical or chiral degrees of freedom as for Dirac fermions in 2 or 3 dimensions.  There are no ``separate" spin degrees of freedom that can be used to construct Heisenberg antiferromagnets and the spin systems required for the spin stripes. The way to go is to further enumerate {\em non-relativistic} holographic setups, so that one can contemplate holographic constructions involving spin order of the kind that is ubiquitous in condensed matter systems. This theme of intertwined order illustrates 
in an effective manner that there is still much realistic quantum matter physics to be explored in the holography laboratory.

{\em Acknowledgements.}\;\;\;
We thank A. Krikun and S.A. Kivelson for helpful discussions. RGC was supported by NNSFC  with Grant No.11375247, No.11435006, and No. 11647601 and by the Key Research Program of Frontier Sciences of CAS.

\onecolumngrid
\newpage

\begin{center}
\textbf{\large Supplemental Material: Optical conductivity in the homogeneous case and breaking of parity}
\end{center}
\setcounter{equation}{0}
\setcounter{figure}{0}
\setcounter{table}{0}
\setcounter{page}{1}
\makeatletter
\renewcommand{\theequation}{S\arabic{equation}}
\renewcommand{\thefigure}{S\arabic{figure}}
\renewcommand{\bibnumfmt}[1]{[S#1]}
\renewcommand{\citenumfont}[1]{S#1}

We show here that the holographic theory of the main text yields indeed a dual description of a superconductor by calculating the optical conductivity. A good starting point is to consider the homogeneous case where all background fields only depend on the radial direction $z$. We take the probe limit where the back-reaction of the matter sources to the background can be ignored. It has the advantage of avoiding the undesired infinite DC conductivity typical of translationally invariant charged media, due to the fact that the conserved momentum overlaps with the current operator leading to delta functions in the conductivity. In the probe limit there is no overlap between the momentum and the current. Therefore, the observation of the delta function at zero frequency is an unambiguous signature of the U(1) symmetry breaking.

The background is taken to be the AdS Schwarzschild black hole:
\begin{eqnarray}\label{schw}
ds^2=\frac{1}{z^2}\left[- f(z)dt^2+\frac{1}{f(z)}dz^2+(dx^2+dy^2)\right]\,, \quad f(z)=1-z^3\,.
\end{eqnarray}
To illustrate matters, let us consider the model with 
\begin{eqnarray}
Z=1\,,\quad \mathcal{F}=\frac{1}{2}\chi^2\,,\quad V=-\chi^2\,,\quad \vartheta=\frac{n}{2} \chi\,,
\end{eqnarray}
which is fully representative for the physics of interest. In general, we consider a theory in which $\vartheta$ is an odd function of $\chi$ while others are even functions, so that the theory preserves parity and time-reversal invariance. We seek for solutions of the Ansatz
\begin{eqnarray}
\chi=\chi(z)\,,\quad A=\phi(z)\,dt\,,
\end{eqnarray}
while the other fields are trivial. The equations of motion become
\begin{eqnarray}
\phi''-\frac{\chi^2}{z^2 f} \phi&=&0\,,\\
\chi''+\left(\frac{f'}{f}-\frac{2}{z}\right)\chi'+\frac{1}{f}\left(\frac{\phi^2}{f}+\frac{2}{z^2}\right)\chi&=&0\,.
\end{eqnarray}
In the normal phase $\chi=0$ and $\phi$ is then solved by $\phi=\mu(1-z)$ with $\mu$ being the chemical potential. The symmetry-broken phase in the boundary characterized by non-trivial scalar hair in the bulk is obtained by numerically solving the system with source free boundary condition in the UV. We adopt the standard choice by identifying the leading asymptotic term as the source term and the subleading term the response $\langle O\rangle$~\cite{Hartnoll:2008}. A broken phase develops spontaneously below $T_c\approx 0.059\mu$, being thermodynamically favored over the normal phase.

Now we turn to study the optical conductivity in this symmetry broken phase, by turning on U(1) gauge field perturbations $a_x$ and $a_y$ with harmonic time dependent $e^{-i\omega t}$. This results in
 two coupled equations of motion:
\begin{eqnarray}
a_x''+\frac{f'}{f}a_x'-\left(\frac{\chi^2}{z^2f}-\frac{\omega^2}{f^2}\right)a_x+\frac{4\,i\,n\,\omega\,\chi'}{f}a_y&=&0\,,\label{axeom}\\
a_y''+\frac{f'}{f}a_y'-\left(\frac{\chi^2}{z^2f}-\frac{\omega^2}{f^2}\right)a_y-\frac{4\,i\,n\,\omega\,\chi'}{f}a_x&=&0\,.\label{ayeom}
\end{eqnarray}
Notice that the CS terms enter into the perturbation equations. As the causal behavior is related to the retarded two-point function of the current, we impose  ingoing boundary conditions at the horizon. The asymptotic behavior of the U(1) field near the UV boundary $z\rightarrow 0$ is seen to be
\begin{eqnarray}
a_x=a_x^s+a_x^v\, z+\cdots\,,\quad a_y=a_y^s+a_y^v\, z+\cdots\,.
\end{eqnarray}
According to the holographic dictionary, the leading- and the subleading terms are the dual source and expectation value of the current, respectively.
The conductivity using the boundary data can  subsequently be obtained via
\begin{equation}
\begin{pmatrix}
      a_x^v    \\
      a_y^v
\end{pmatrix}
=\begin{pmatrix}
     \sigma_{xx} & \sigma_{xy}   \\
      \sigma_{yx}&  \sigma_{yy}
\end{pmatrix}
\begin{pmatrix}
      i\, \omega\, a_x^s    \\
      i \,\omega\, a_y^s
\end{pmatrix}.
\end{equation}
The rotational invariance of the present setup implies that $ \sigma_{xx}= \sigma_{yy}$ and $ \sigma_{xy}=- \sigma_{yx}$.

\begin{figure}[ht!]
\begin{center}
\subfigure[]{\includegraphics[scale=0.9]{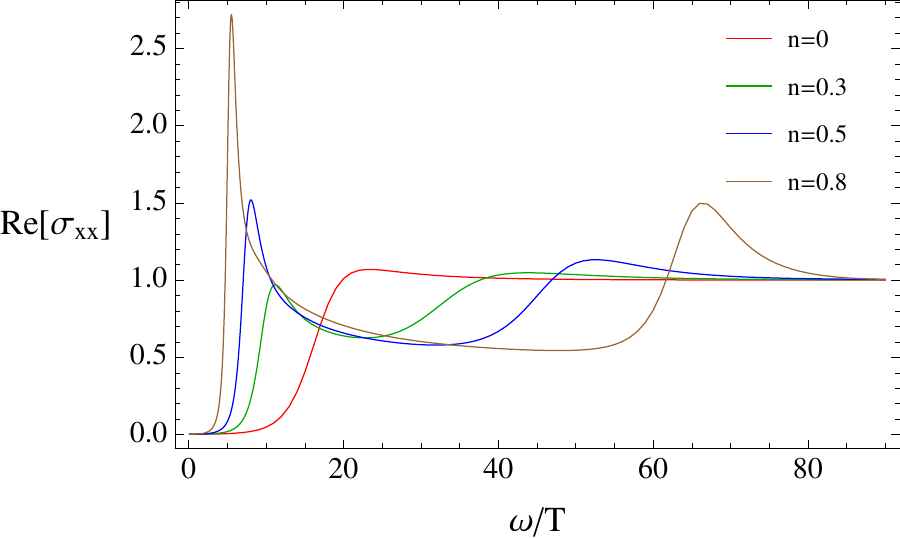}}\quad\quad
\subfigure[]{\includegraphics[scale=0.9]{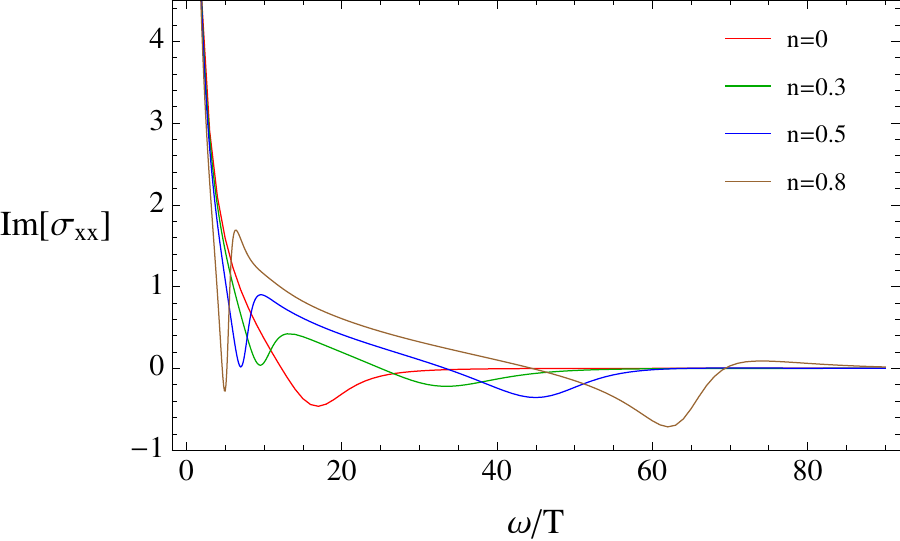}}
\caption{ Panel (a) shows the real part of the longitudinal conductivity for the broken phase at $T/T_c\approx0.52$ for different choice of CS coupling constant $n$. Panel (b) gives the corresponding imaginary part of $\sigma_{xx}$. The red curve is the case without CS term, while others have non-trivial contribution from CS coupling. There is a pole in the imaginary part at $\omega=0$, therefore the real part should have a delta function there.}
\label{fig:sigxx}
\end{center}
\end{figure}

We have calculated the electrical conductivity for different values of CS coupling constant $n$ and for different temperatures for the black holes encoding for the ordered phase. 
In  Fig.\,\ref{fig:sigxx} the conductivity is shown as function of frequency $\omega$. The real part of the conductivity contains a delta function at zero frequency $\omega=0$ as follows from the behavior of the imaginary part Im$[\sigma_{xx}(\omega)]$ exhibiting the characteristic $1/\omega$ dependence.
On the other hand, the conductivity for the unbroken phase is finite. In our present setup $\sigma_{xx}=1$ as a consequence of the electric-magnetic duality~\cite{Herzog:2007ij}.  It follows that the delta function at zero frequency is uniquely associated with the spontaneously U(1) symmetry breaking. It will be interesting to compute the optical conductivity in the uni-directional and  the fully crystallised phases. This is a quite involved numerical affair  and we hope to report results in a near future. However, the simplified procedure outlined in the above demonstrates unambiguously that 
our ``holographic pair density wave" is a genuine superconductor.    

Finally, it should be pointed out that in the broken phase with non-trivial $\chi$ there is a spontaneously breaking of parity. This can be easily understood from the perturbation equations~\eqref{axeom} and~\eqref{ayeom}. The parity operation in the dual boundary field theory can be defined as the reversal of one of the spatial directions~\cite{Gauntlett:2009}, say $x$, which transforms $a_x\rightarrow -a_x$ while keeping $a_y$ unchanged. On the other hand, the frequency $\omega$ changes its sign under the time reversal. From the last terms originating from the CS coupling $\vartheta(\chi) F\wedge F$ in~\eqref{axeom} and~\eqref{ayeom}, one finds that non-trivial configurations of $\chi$ break parity and time-reversal invariance of the boundary field theory. In contrast, there is no parity violation for the normal phase with vanishing $\chi$. In some sense the VEV of the operator dual to $\chi$ plays the role of order parameter for parity breaking. For the the uni-directional and the fully crystallised phases of the main text, there are spontaneous currents as a consequence of spontaneous parity symmetry breaking. Therefore, the theory of the main text provides a dual description of an inhomogeneous superconducting phase intertwined with charge, current and parity orders. Although there are some other ways to spontaneously break translational symmetry in holography, {\em e.g.}~\cite{Donos:2013}, they preserve the parity and usually need to fine tune the model parameters. To incorporate parity breaking effect that has been sharply observed in the cuprates, a natural and much easier way is to invoke the topological (theta and CS) terms.


\begin{thebibliography}{99}

\bibitem{naturerev15} B.~Keimer, S.~A.~Kivelson, M.~R.~Norman, S.~Uchida and J.~Zaanen, Nature {\bf 518}, 179 (2015). 

\bibitem{fradkinPDW} E.~Fradkin, S.~A.~Kivelson and J.~M.~Tranquada, Rev. Mod. Phys. {\bf 87}, 457 (2015). 

\bibitem{currentsfirst} B.~Fauqu\'e {\em et al.}, Phys. Rev. Lett.  {\bf 96}, 197001 (2006). 

\bibitem{currentsGreven} Y.~Li  {\em et al.}, Nature {\bf 455}, 372 (2008).

\bibitem{currenttflucGreven} Y.~Li  {\em et al.}, Nature {\bf 468}, 283 (2010).


\bibitem{parityodd} L.~Zhao {\em et al.}, Nature Physics {\bf 13}, 250 (2017).

\bibitem{tranquadaPDW07} Q.~Li, M.~H\"ucker, G.~D.~Gu, A.~M.~Tsvelik and J.~M.~Tranquada, Phys. Rev. Lett. {\bf 99}, 067001 (2007). 


\bibitem{Rajasekaran:2017} S.~Rajasekaran {\em et al.}, arXiv:1705.06112.

\bibitem{DavisPDW16} M.~H.~Hamidian {\em et al.}, Nature {\bf 532}, 343 (2016). 

\bibitem{Devstripes} E.~W.~Huang {\em et al.}, arXiv:1612.05211.

\bibitem{Simonsnum} B.~X.~Zheng {\em et al.}, arXiv:1701.00054.

\bibitem{thebook15} J.~Zaanen, Y.~W.~Sun, Y.~Liu and K.~Schalm, ``Holographic duality in condensed matter physics" (Cambridge Univ. Press, 2015). 

\bibitem{Hartnoll:2016apf} 
  S.~A.~Hartnoll, A.~Lucas and S.~Sachdev,
  arXiv:1612.07324 [hep-th].

\bibitem{Hartnoll:2008vx} 
  S.~A.~Hartnoll, C.~P.~Herzog and G.~T.~Horowitz,
  Phys.\ Rev.\ Lett.\  {\bf 101}, 031601 (2008).
  
\bibitem{She:2011cm} 
  J.~H.~She, {\em et al.},
  Phys.\ Rev.\ B {\bf 84}, 144527 (2011).

\bibitem{Nakamura:2009tf} 
  S.~Nakamura, H.~Ooguri and C.~S.~Park,
  Phys.\ Rev.\ D {\bf 81}, 044018 (2010).
  
\bibitem{Donos:2011bh} 
  A.~Donos and J.~P.~Gauntlett,
  JHEP {\bf 1108}, 140 (2011).
  
\bibitem{Withers:2014sja} 
  B.~Withers,
  JHEP {\bf 1409}, 102 (2014).
  
\bibitem{Nayak} C.~Nayak, Phys. Rev.  B {\bf 62}, 4880 (2000).

\bibitem{Chakravarty} S.~Chakravarty, R.~B.~Laughlin, D.~K.~Morr and C.~Nayak, Phys. Rev. B {\bf 63}, 094503 (2001).

\bibitem{affleck} I.~Affleck and J.~B.~Marston, Phys. Rev. B {\bf 37}, 3774(R) (1988).

\bibitem{Kotliar} G.~Kotliar, Phys. Rev. B {\bf 37}, 3664 (1988).  
  
\bibitem{Gauntlett:2009bh} 
  J.~P.~Gauntlett, J.~Sonner and T.~Wiseman,
  JHEP {\bf 1002}, 060 (2010).
  
\bibitem{Rozali:2012es} 
  M.~Rozali, D.~Smyth, E.~Sorkin and J.~B.~Stang,
  Phys.\ Rev.\ Lett.\  {\bf 110}, no. 20, 201603 (2013).
  
\bibitem{Withers:2013kva} 
  B.~Withers,
  arXiv:1304.2011 [hep-th].
  
\bibitem{Donos:2013wia} 
  A.~Donos,
  JHEP {\bf 1305}, 059 (2013).
  
  
\bibitem{Donos:2015eew} 
  A.~Donos and J.~P.~Gauntlett,
  JHEP {\bf 1603}, 148 (2016).
 


\bibitem{Donos:2013gda} 
  A.~Donos and J.~P.~Gauntlett,
  Phys.\ Rev.\ D {\bf 87}, no. 12, 126008 (2013)


\bibitem{Franco:2009yz} 
  S.~Franco, A.~Garcia-Garcia and D.~Rodriguez-Gomez,
  JHEP {\bf 1004}, 092 (2010).
  
\bibitem{Aprile:2009ai} 
  F.~Aprile and J.~G.~Russo,
  Phys.\ Rev.\ D {\bf 81}, 026009 (2010).
  
\bibitem{Cai:2012es} 
  R.~G.~Cai, S.~He, L.~Li and L.~F.~Li,
  JHEP {\bf 1210}, 107 (2012).
  
\bibitem{Cremonini:2016rbd} 
  S.~Cremonini, L.~Li and J.~Ren,
  Phys.\ Rev.\ D {\bf 95}, no. 4, 041901(R) (2017).
  
\bibitem{Gouteraux:2016arz} 
  B.~GoutŽraux and V.~L.~Martin,
  JHEP {\bf 1705}, 005 (2017).


\bibitem{addsm}
See Supplemental Material for the discussion about the optical conductivity and breaking of parity supporting statements made in the Letter, which includes Refs.~\cite{Hartnoll:2008vx,Gauntlett:2009bh,Donos:2013gda,Herzog:2007}

\bibitem{Herzog:2007} 
  C.~P.~Herzog, P.~Kovtun, S.~Sachdev and D.~T.~Son,
  Phys.\ Rev.\ D {\bf 75}, 085020 (2007).

\bibitem{Cai:1996eg} 
  R.~G.~Cai and Y.~Z.~Zhang,
  Phys.\ Rev.\ D {\bf 54}, 4891 (1996).
  
\bibitem{CLWZ}
R.~G.~Cai, L.~Li, Y.~Q. Wang and J.~Zaanen, to appear.

\bibitem{Headrick:2009pv} 
  M.~Headrick, S.~Kitchen and T.~Wiseman,
  Class.\ Quant.\ Grav.\  {\bf 27}, 035002 (2010).

\bibitem{Zachar} O.~Zachar, S.~A.~Kivelson and V.~J.~Emery, Phys. Rev. B {\bf 57}, 1422 (1998). 



\bibitem{Varma} C.~M.~Varma, Nature {\bf 468}, 184 (2010).
 
\bibitem{Shekhter} A.~Shekhter {\em et al.}, Nature {\bf 498}, 75 (2013). 


\bibitem{Andrade:2017leb} 
  T.~Andrade and A.~Krikun,
  JHEP {\bf 1703}, 168 (2017).


\bibitem{ZaGu} J.~Zaanen and O.~Gunnarsson, Phys. Rev. B {\bf 40}, 7391(R) (1989).

\bibitem{mesarosSTS} A.~Mesaros {\em et al.}, PNAS 113, 12661 (2016). 



\end{thebibliography}

\begin{thebibliography}{11}

\bibitem{Hartnoll:2008} 
  S.~A.~Hartnoll, C.~P.~Herzog and G.~T.~Horowitz,
  ``Building a Holographic Superconductor,''
  Phys.\ Rev.\ Lett.\  {\bf 101}, 031601 (2008).
 [arXiv:0803.3295 [hep-th]].

\bibitem{Herzog:2007ij} 
  C.~P.~Herzog, P.~Kovtun, S.~Sachdev and D.~T.~Son,
  ``Quantum critical transport, duality, and M-theory,''
  Phys.\ Rev.\ D {\bf 75}, 085020 (2007).
  [hep-th/0701036].

\bibitem{Gauntlett:2009} 
  J.~P.~Gauntlett, J.~Sonner and T.~Wiseman,
  ``Quantum Criticality and Holographic Superconductors in M-theory,''
  JHEP {\bf 1002}, 060 (2010)
  [arXiv:0912.0512 [hep-th]].
  
\bibitem{Donos:2013} 
  A.~Donos and J.~P.~Gauntlett,
  ``Holographic charge density waves,''
  Phys.\ Rev.\ D {\bf 87}, no. 12, 126008 (2013)
 [arXiv:1303.4398 [hep-th]].
  

\end{thebibliography}
\end{document}